\documentstyle[tighten,eqsecnum,prl,aps]{revtex}
\begin{document}
%
% ***  definition  ***
%
\def\eps{\varepsilon}
\def\hsp{\hspace*{-0.6cm}}
\draft
\title
{
Anisotropic pseudogap in the half-filling $2$-$d$ Hubbard model
at finite $T$
}
\author{
Taiichiro Saikawa and Alvaro Ferraz
}
\address{
Laborat\'{o}rio de Supercondutividade,\\
Centro Internacional de F\'{\i}sica da Mat\'{e}ria Condensada,\\
Universidade de Bras\'{\i}lia,\\
CEP 70919-970 Bras\'{\i}lia-DF, Brazil
}

\date{October 23, 1998}
\maketitle
\begin{abstract}
We have studied the pseudogap formation in the single-particle spectra
of the half-filling two-dimensional Hubbard model.
Using a Green's function with the one-loop self-energy correction
of the spin and charge fluctuations, we have numerically calculated
the self-energy, the spectral function, and the density of states
in the weak-coupling regime at finite temperature.
Pseudogap formations have been observed in both the density of states
and the spectral function at the Fermi level.
The pseudogap in the spectral function is explained by
the non-Fermi-liquid-like nature of the self-energy.
The anomalous behavior in the self-energy is caused by both
the strong antiferromagnetic spin fluctuation
and the nesting condition on the non-interacting Fermi surface.
In the present approximation, we find a logarithmic singularity
in the integrand of the imaginary part of the self-energy.
Concerning the energy dependence of the spectral function and
the self-energy, two theorems are proved.
They give a necessary condition in the self-energy
to produce the pseudogap at the Fermi level.
The pseudogap in the spectral function is highly momentum dependent
on the Fermi surface.
It opens initially in the $(\pm \pi,0)$, $(0,\pm \pi)$ regions
as the normal state pseudogap observed in the high-$T_c$ superconductors
and if the interaction is increased, it spreads to other
Fermi surface sectors.
The anisotropy of the pseudogap is produced by the low-energy enhancement
of the spin excitation around ${\bf Q}=(\pi,\pi)$ and
the flatness of the band dispersion around the saddle point.
\end{abstract}


%%%%%%%%%%%%%%%%%%%%%%%%%%%%%%%%%%%%%%%%%%%%%%%%%%%%%%%%%%%%%%%%%%%%%%%
\section{Introduction}

The recent series of angle-resolved-photoemission experiments
in high-$T_c$ superconductors
have brought a large number of new insights concerning
the low energy single-particle states of these materials.
The experimental observations include a rather wide
flat-band region around $(\pi,0)$ in the effective band
dispersion\cite{GOFRON94,KING94,ZXSDES95}
and also the anisotropy of the normal state pseudogap
\cite{MARSHALL96,LOESER96,DING96,ZXSHEN97,NORMAN98,NORMAN98A}
which is consistent with the $d$-wave symmetry superconducting gap.
Many theoretical approaches and computer simulations have
been performed to investigate the pseudogap.
Naturally, our ultimate objective
is to explain the existing experimental data.
However, the main purpose of this work
is not to give a direct explanation of the experiments,
but rather, to study the pseudogap formation in the Hubbard model
which is often used to describe the high-$T_c$ materials.

To avoid confusion, we should make clear what kinds of pseudogap
we are dealing with in this paper.
In reality, in the literature this term has been used to represent
several different features.  What it is common in all cases
is the fact that
the pseudogap indicates a suppression or disappearance of
the spectral intensity at the Fermi level.
We can list the following three different cases.
The first one is a disappearance of the spectral intensity
due to the downward shift of the band dispersion around $(\pi,0)$
region\cite{PI0}.
For example, some of the quantum Monte Carlo (QMC) simulations show
such behavior in the strong coupling regime and
at very low dopings\cite{BULUT94c,PREUSS95,PREUSS97}.
This phenomenon is quite interesting since it looks as
if it violates the Luttinger theorem on the Fermi surface.
The second case is an effective pseudogap in the spectral function
between a strong quasiparticle peak and a weak satellite peak.
It is possible to find this situation even in a normal Fermi liquid.
The third one is a suppression of the single-particle peak
at the Fermi level with a two peak structure in its place.
The pseudogap of our interest is this latter case.  It should be
clearly distinguishable from the second case.
One physical origin of this kind of pseudogap is 
of course the fact that it is a precursor
of the spin density wave (SDW) transition
which takes place in the $2$-$d$ Hubbard model
for zero doping at $T=0$.
We are aware of some models which take into account
precursor effects of the superconducting transition
to explain the pseudogap behavior observed in the normal phase
of high-$T_c$ materials\cite{JANKO97,ENGEL97,VILK97z,KORNI98}.
We don't consider the superconducting transition effects
in this paper.

In our work, we provide an explanation of the origin
of the pseudogap formation in the $2$-$d$ Hubbard model,
at half-filling, in the weak coupling regime.
To calculate the spectrum we use a single-particle Green's function
which has a paramagnon-like
one-loop self-energy correction for both charge and spin channels.
At first we show the results of our numerical calculations
of the density of states (DOS), the spectral function,
and the self-energy.
The analysis of the real and imaginary parts of the self-energy plays
a crucial role in the understanding of the microscopic origin of
the possible structures of the spectra.  Following this, we discuss
the detailed origin of the pseudogap formation within our formulation.

We found pseudogap formations
at the Fermi level in both the spectral function and the DOS.
The pseudogap in the spectral function is explained by
the non-Fermi-liquid behavior of the self-energy, i.e.,
its imaginary part has a negative peak and its real part has
a positive slope.
We show that both the strong antiferromagnetic spin fluctuation
and the nesting condition
on the non-interacting Fermi surface are the origins of
the anomalous behaviors in the self-energy.
We also generally argue the relation between the spectral function
and the self-energy at the Fermi level.
We show, for example, that if the spectral function has a pseudogap,
the real part of the self-energy has a positive slope
at the Fermi level.
An auxiliary relation is used to present the argument.
This relation derived from the Kramers-Kronig transformation
is proved in the appendix.

The numerically obtained pseudogap shows a strong momentum
dependence.  For a certain choice of the parameters, the pseudogap
appears around $(\pi,0)$ but not at $(\pi/2,\pi/2)$.
We show that the anisotropic pseudogap comes from
the low-energy enhancement of the spin excitation around
${\bf Q}=(\pi,\pi)$ and the flatness of the band dispersion.

%%%%%%%%%%%%%%%%%%%%%%%%%%%%%%%%%%%%%%%%%%%%%%%%%%%%%%%%%%%%%%%%%%%%%%%%%%
\section{Model}

In the momentum representation the Hubbard model,
with a nearest-neighbor hopping, on a two-dimensional square lattice
can be written as
\begin{equation}
   H = \sum_{{\bf k}, \sigma} (\eps_{\bf k} - \mu)
          a^{\dagger}_{{\bf k} \sigma} a_{{\bf k} \sigma}
      +\frac{U}{2} \sum_{{\bf k}, {\bf p}_1, {\bf p}_2, \sigma }
                 a^{\dagger}_{{\bf p}_1+{\bf k}  \sigma}
                 a^{\dagger}_{{\bf p}_2-{\bf k} -\sigma}
                 a_{{\bf p}_2 -\sigma}
                 a_{{\bf p}_1  \sigma}
\end{equation}
with $\eps_{\bf k} = -2 t ( \cos k_x + \cos k_y )$,
$\sigma$ representing the spin index and with $U$ being
the coupling constant of the on-site
Coulomb repulsion between two electrons with opposite spins.
We redefine the chemical potential $\mu$ by adding
the Hartree-Fock term and this is chosen as the origin of the energy
for the half-filling case.

We next introduce the single-particle Green's function
at finite temperature in the paramagnetic regime.
In general, we can write the full single-particle
Green's function as
\begin{equation}
G({\bf k},\omega_n)^{-1}
 = G_0({\bf k},\omega_n)^{-1}
   -\Sigma({\bf k},\omega_n)
\label{DEQF}
\end{equation}
where $G_0({\bf k}, \omega_n)^{-1} = i\omega_n - \eps_{\bf k}$
and $\Sigma({\bf k},\omega_n)$ is the proper self-energy.
Here, in the weak-coupling regime, we consider the one-loop
self-energy correction to the Green's function.
The formulation we use is based on
the standard perturbation expansion over $U/t$.
These one-loop corrections are constructed with the free
(mean-field) fermion line and
the corresponding charge or spin fluctuation lines.
The charge and spin fluctuations are described in terms of
the susceptibilities calculated within the random phase approximation
(RPA).
We neglect higher-order contributions which appear
in the diagrammatic series expansion.
This kind of approximation has already been used
in paramagnon theory\cite{PARAMAG93}.
Thus, the one-loop self-energy is simply
\begin{eqnarray}
\Sigma({\bf k}, \omega_n) &=&
   U^2 T\sum_{\nu_m} \sum_{\bf q}
    \left[
    \frac{1}{2}\chi_c({\bf q},\nu_m)
    \right.
\nonumber \\
& & \left.
    +\frac{3}{2}\chi_s({\bf q},\nu_m)
    -\chi_0({\bf q},\nu_m)
    \right] G_0({\bf k}+{\bf q},\omega_n+\nu_m)
    \label{SIGPARA}
\end{eqnarray}
where we have defined the charge susceptibility as
$\chi_c({\bf q},\nu_m) =
    \chi_0({\bf q},\nu_m)/[1+U\chi_0({\bf q},\nu_m)] $
and the spin susceptibility as
$\chi_s({\bf q},\nu_m) =
    \chi_0({\bf q},\nu_m)/[1-U\chi_0({\bf q},\nu_m)] $.
To avoid the double counting of the second-order self-energy diagram,
we have subtracted $\chi_0({\bf q},\nu_m)$ in the square brackets
in $\Sigma({\bf k}, \omega_n)$.
Here $\chi_0$ is given by
\begin{equation}
\chi_0({\bf q}, \nu_m) = {\sum_{\bf k}}
   \frac{f(\eps_{{\bf k}+{\bf q}})-f(\eps_{\bf k})}
   {i\nu_m -(\eps_{{\bf k}+{\bf q}} - \eps_{\bf k})}
   \label{CHI00}
\end{equation}
with $\nu_m= 2 m \pi T$
and $f(x)$ is the Fermi distribution function defined
by $f(x) = 1/[\exp(x/T) + 1]$.

%%%%%%%%%%%%%%%%%%%%%%%%%%%%%%%%%%%%%%%%%%%%%%%%%%%%%%%%%%%%%%%%%%%%%%%%%%
\section{Numerical Results}

To calculate the spectra obtained from the above Green's function,
we made the analytic continuation of the energy variables
of the self-energy and the Green's function.
In this way we obtained the retarded self-energy and
the retarded Green's function.
As a result, the imaginary part of the self-energy becomes,
\begin{eqnarray}
   {\rm Im}\Sigma^R({\bf k}, \omega)
 &=& -U^2 {\sum_{\bf q}}
     \left[
   \frac{3}{2}{\rm Im}\chi_s^R({\bf q},\eps_{{\bf k}+{\bf q}}-\omega)
  +\frac{1}{2}{\rm Im}\chi_c^R({\bf q},\eps_{{\bf k}+{\bf q}}-\omega)
  -{\rm Im}\chi_0^R({\bf q},\eps_{{\bf k}+{\bf q}}-\omega)
     \right]
\nonumber \\
 & & \times
     \left[
       f(\eps_{{\bf k}+{\bf q}}) + n(\eps_{{\bf k}+{\bf q}}-\omega)
     \right]
\end{eqnarray}
where $n(x)$ is the Bose distribution function defined as
$n(x) = 1/[\exp(x/T)-1]$.  `R' denotes the retarded function.
The corresponding real part can be obtained from the imaginary part
by means of the Kramers-Kronig transformation.
Using those schemes, we have numerically calculated
the susceptibilities, the self-energy, and the Green's function.
The momentum summation in ${\rm Im}\chi_0({\bf q},\nu)$ was
reduced to a contour integral in the Brillouin zone and roughly $4000$
line-elements have been numerically summed up along the contour.
The momentum summation in the imaginary part of the self-energy
and the DOS have been done on a $120 \times 120$ mesh
in the Brillouin zone.
All parameters used in the calculation are within the Stoner
instability condition in RPA.

%%%%%%%%%%%%%%%%%%%%%%%%%%%%%%%%%%%%%%%%%%%%%%%%%%%%%%%%%%%%%%%%%%%%%%%%%%
\subsection{Density of States}

In Fig.\ \ref{DOS1} we show the $U$ dependence of
the DOS $N(\omega) = \sum_{\bf k} A({\bf k},\omega)$ for $T/t=0.5$
where the spectral function $A({\bf k}, \omega)$ is defined by
\begin{eqnarray}
A({\bf k},\omega) &=& -\frac{1}{\pi}{\rm Im}G^R({\bf k},\omega)
\nonumber \\
 &=& -\frac{1}{\pi}
 \frac{ {\rm Im}\Sigma^R({\bf k},\omega) }
 {[ \omega-\eps_{\bf k}-{\rm Re}\Sigma^R({\bf k},\omega)]^2
 +[ {\rm Im}\Sigma^R({\bf k},\omega)]^2}.
 \label{AKW}
\end{eqnarray}
For $U/t=1.5$, the overall structure still keeps the shape
similar to the non-interacting case with
a clear single peak on the Fermi level.
For $U/t=2.5$, a weak pseudogap can be seen, and finally,
for $U/t=3.0$, a rather wide pseudogap can be observed.
The high-energy band tails are developed as $U/t$ becomes large.

Figure \ref{DOS2} shows the temperature dependence of the pseudogap
formation in the density of states for $U/t=2.0$.
The pseudogap develops as we lower the temperature.
Using the Stoner condition, we numerically found that
$T/t=0.21$ is the SDW transition temperature for $U/t=2.0$.
The temperature effect appears only around the Fermi level
and the overall high-energy structure is left almost unchanged.

A similar calculation of the DOS was done
in the work by Kampf and Schrieffer\cite{KAMPF90a}.
They have also obtained the pseudogap formation near
the Fermi level $\omega=0$ for a weakly doped Hubbard model
at zero temperature.  In their calculation they took into account
only the spin fluctuation effects in an one-loop self-energy
similar to ours.  Since the spin fluctuation effect becomes
much stronger than the charge contribution near the SDW transition,
the similarity of our calculation and theirs is natural.
Pseudogap openings in the DOS in the weak-coupling regime
has been observed in a QMC simulation\cite{CREF95}.
They obtained pseudogaps
for $U/t=4.0$ and $T/t=0.2$ for several lattice sizes
from $8 \times 8$ to $12 \times 12$.
A similar behavior in the DOS obtained by the fluctuation exchange
(FLEX) approximation\cite{DEISZ96}.
They obtained a very narrow and deep pseudogap
in the DOS for $U/t=1.57$ and $T/t=0.05$ at half-filling.
All of those results including ours are at least qualitatively
in good agreement with each other.
The DOS for $U/t=3.0$ in Fig.\ \ref{DOS1}
has a rather large pseudogap.  This may be due however to
our simple approximation.

%%%%%%%%%%%%%%%%%%%%%%%%%%%%%%%%%%%%%%%%%%%%%%%%%%%%%%%%%%%%%%%%%%%%%%%%%%
\subsection{Spectral function and the self-energy}

The existence of the pseudogap in the DOS enables us to guess
the decrease of the intensity of the spectral function at $\omega=0$.
In Fig.\ \ref{SPCFIG}, we plot the temperature dependence
of the spectral function $A({\bf k}_F,\omega)$
at ${\bf k}_F=(\pi,0)$ together with
the real and imaginary parts of the self-energy.
We have also calculated $A({\bf k},\omega)$ for other momenta
and our results are a strong indication of the
anisotropy of the pseudogap.
We will discuss it in more detail in a later section.

As we have expected, Fig.\ \ref{SPCFIG} (a) shows
the weight decrease in the spectral function
as the pseudogap develops in the DOS shown in Fig.\ \ref{DOS2}.
The self-energy shows a specific behavior.
The real part (Fig.\ \ref{SPCFIG} (b)) has a positive slope and
the imaginary part (Fig.\ \ref{SPCFIG} (c)) shows a negative peak.
This tendency becomes stronger as we decrease
the temperature.
Moreover, the Green's function has three poles
which correspond to the solutions of
$\omega-\eps_{{\bf k}_F} = {\rm Re}\Sigma^R({\bf k}_F,\omega)$
for $T/t=0.22$ as shown in Fig.\ \ref{SPCFIG} (b).
Two of the poles are associated with two peaks of $A({\bf k}_F,\omega)$
in Fig.\ \ref{SPCFIG} (a)
and the other pole is linked to the pseudogap.
Those features in the spectral function and
the self-energy clearly indicate the destruction of the Fermi-liquid
quasiparticle states.
We will argue in more detail on the relation between the pseudogap and
the anomalous behavior in the self-energy in the next section.

This same trend in the spectral function has already been
shown in the so-called two-particle self-consistent (TPSC)
approach by Vilk and Tremblay\cite{VILK96,VILK97a}.
In the FLEX approach\cite{DEISZ96}, however, there is no pseudogap
in the spectral function although their self-energy
shows a similar non-Fermi-liquid behavior that both our result
and TPSC approach have shown.
The detailed analysis and comparisons between those approaches
can be seen in Ref.\ \onlinecite{VILK97a}.

For the weak coupling regime at half-filling,
the finite temperature pseudogap formation in $A({\bf k},\omega)$
in a QMC simulation\cite{WHITE92} was attributed to a finite
size effect.
Making a different analysis of the QMC simulation data, the pseudogap
formation was confirmed by Creffield et al\cite{CREF95}.
The latter authors have used a singular-value-decomposition method
instead of the maximum-entropy method
to obtain the spectral function from the calculated finite temperature
Green's function.  They obtained a clear pseudogap opening
even with a $12 \times 12$ system size.

%%%%%%%%%%%%%%%%%%%%%%%%%%%%%%%%%%%%%%%%%%%%%%%%%%%%%%%%%%%%%%%%%%%%%%%%%%
\section{Pseudogap and
the non-Fermi-liquid behavior in the self-energy}

In this section we discuss the origin of the non-Fermi-liquid
behavior in the imaginary part of the self-energy.
This section is organized in three subsections.
Firstly we analyze the imaginary part of the self-energy
within the present paramagnon theory.
A log singularity is found in the integrand of the imaginary part
of the self-energy.
Next, we examine a possibility of having the same tendency of
the imaginary part of the self-energy using a model susceptibility
which does not produce the log singularity.
Finally, we discuss a general relation between the self-energy
and the spectral function around the Fermi level.
The argument in the final subsection
does not depend on the models used and on the origin
of the pseudogap.

%%%%%%%%%%%%%%%%%%%%%%%%%%%%%%%%%%%%%%%%%%%%%%%%%%%%%%%%%%%%%%%%%%%%%%%%%%
\subsection{Log singularity in the integrand of the imaginary part
of the self-energy}

We can now explain the origin of the non-Fermi-liquid negative peak
behavior in ${\rm Im}\Sigma({\bf k},\omega)$ in the present formulation.
The point is how to take into account the strong enhancement
of the spin fluctuation
around ${\bf q}=(\pi, \pi) \equiv {\bf Q}$.
Let's consider the integrand of the spin
component of the self-energy at ${\bf k}={\bf k}_F$.
The integrand is 
\begin{equation}
I_{{\bf k}_F}({\bf q},\omega)
=-\frac{3}{2}U^2
{\rm Im}\chi_s^R({\bf q},\eps_{{\bf k}_F+{\bf q}}-\omega)
\left[
f(\eps_{{\bf k}_F+{\bf q}}) + n(\eps_{{\bf k}_F+{\bf q}}-\omega)
\right].
\end{equation}

Considering the strong low-energy enhancement of
${\rm Im}\chi_s({\bf Q},\nu)$, we suppose that the major contribution
to the imaginary part of the self-energy comes from
the integrand at ${\bf q}={\bf Q}$.
By setting ${\bf q}={\bf Q}$, the spin fluctuation spectrum
can be written as
\begin{equation}
{\rm Im}\chi_s^R({\bf Q}, \eps_{{\bf k}_F+{\bf Q}}-\omega) =
\frac{{\rm Im}\chi_0^R({\bf Q},-\omega)}
{
\left[1- U {\rm Re}\chi_0^R({\bf Q},-\omega) \right]^2
+\left[U {\rm Im}\chi_0^R({\bf Q},-\omega) \right]^2
}
\end{equation}
with
\begin{equation}
{\rm Im}\chi_0^R({\bf Q},\nu) = \frac{\pi}{2}
         \rho_0 \left( \frac{\nu}{2} \right)
         \tanh \left(\frac{\nu}{4T}\right)
  \label{IMX0}
\end{equation}
and
\begin{equation}
{\rm Re}\chi_0^R({\bf Q},\nu) = \frac{P}{\pi} \int d\nu'
           \frac{{\rm Im}\chi_0^R({\bf Q},\nu')}{\nu'-\nu}.
\end{equation}
Here $\rho_0(x) = (1/(2\pi^2t))
{\rm K}\left( \sqrt{1-[x/(4t)]^2} \right)$ is
the density of states
for the non-interacting band $\eps_{\bf k}$ where ${\rm K}(k)$ is
the complete elliptic integral of the first kind.
In the same way, we also have
\begin{equation}
f(\eps_{{\bf k}_F+{\bf Q}})+n(\eps_{{\bf k}_F+{\bf Q}}-\omega) = 
   -\frac{1}{2}\coth \left( \frac{\omega}{2T} \right).
\end{equation}
Note that we have used the nesting condition
$\eps_{{\bf k}_F+{\bf Q}}=-\eps_{{\bf k}_F}=0$ on the Fermi surface.
Then, we obtain
\begin{equation}
I_{{\bf k}_F}({\bf Q},\omega)
=-\frac{3}{2}U^2
\frac{
 -(\pi/4) \rho_0 \left( \omega/2 \right)
  \tanh \left(-\omega/(4T)\right)
  \coth \left( \omega/(2T) \right)
}
{
\left[1- U {\rm Re}\chi_0^R({\bf Q},-\omega) \right]^2
+\left[U {\rm Im}\chi_0^R({\bf Q},-\omega) \right]^2
}.
\label{IKF}
\end{equation}

For $\omega \sim 0$,
by the cancellation of the $\omega/T$ dependence in $\tanh$ and $\coth$,
we see that the numerator always has a log singularity since
\begin{equation}
 -\frac{\pi}{4} \rho_0 \left( \frac{\omega}{2} \right)
  \tanh \left(\frac{-\omega}{4T}\right)
  \coth \left( \frac{\omega}{2T} \right)
\sim \frac{1}{16 \pi t} \ln\left( \frac{32t}{|\omega|} \right).
\end{equation}
By approximating the denominator with its value at $\omega=0$,
the behavior of the integrand for ${\bf q=Q}$ is dominated
by the log singularity,
\begin{equation}
I_{{\bf k}_F}({\bf Q},\omega)
\sim -\frac{3 U^2}{32 \pi t}
\frac{
\ln\left( 32t / |\omega| \right)
}
{
\left[1- U {\rm Re}\chi_0^R({\bf Q},0) \right]^2
}.
\end{equation}
In lowering the temperature or by increasing $U$,
$1-U{\rm Re}\chi_0^R({\bf Q},0)$ becomes smaller and
the contribution from the region around ${\bf Q}=(\pi,\pi)$
dominates the other region's contribution
and it causes the larger peak of ${\rm Im}\Sigma({\bf k}_F,\omega)$
at $\omega=0$.

In Fig.\ \ref{INTGRD} we plot the temperature dependence of
$I_{{\bf k}_F}({\bf Q},\omega)$ for $U/t=2.0$ using Eq.\ (\ref{IKF}).
Fig.\ \ref{INTGRD} (a) shows the low-energy enhancement of
${\rm Im}\chi_s({\bf Q},\nu)$.
The log divergence at $\omega=0$ exists for any finite values of $T$.
$I_{{\bf k}_F}({\bf Q},\omega)$ shows a remarkable enhancement
at the same time as ${\rm Im}\chi_s({\bf Q},\nu)$ has a sharp increase.
Clearly, one can see that
this enhancement of $I_{{\bf k}_F}({\bf Q},\omega)$ around $\omega=0$
causes the non-Fermi-liquid, i.e., the negative peak structure
in the imaginary part of the self-energy.

%%%%%%%%%%%%%%%%%%%%%%%%%%%%%%%%%%%%%%%%%%%%%%%%%%%%%%%%%%%%%%%%%%%%%%%%%%
\subsection{No log-singularity case}
In the last subsection we have found a logarithmic behavior
in the integrand of the imaginary part of the self-energy.
This has a strong influence in producing the non-Fermi-liquid behavior
in the self-energy.
In general, however, the logarithmic van Hove singularity in the DOS
will be smeared to some extent
by any finite interaction between electrons.
One might suppose that the non-Fermi-liquid-like behavior we have
seen in the last section is an artifact of our approximation.
In this subsection, we consider this question by introducing a model
susceptibility which does not possess the log-singularity behavior.
Again we take into account only the spin components and neglect
other channels in the self-energy.

From ${\rm Im}\chi_0^R({\bf Q},\nu)$ given by Eq.\ (\ref{IMX0}),
we see that the logarithmic behavior comes from
the non-interacting density of states $\rho_0(\omega)$.
To consider the case that $\rho_0(\omega)$ does not have
any divergence at $\omega=0$,
we simply assume that $\rho_0(\omega)$ is finite at $\omega=0$.
Then, from Eq.\ (\ref{IMX0}),
we can approximate the imaginary part of $\chi_0({\bf Q},\nu)$ as
${\rm Im}\chi_0^R({\bf Q},\nu) = c \nu$ around $\nu=0$
where $c$ is a positive constant.
Since we would like to
see the low-energy behavior of the integrand of ${\rm Im}\Sigma$,
it is sufficient to approximate the spin susceptibility
around $\nu=0$.
Furthermore, approximating ${\rm Re}\chi_0({\bf Q},\nu)$
by its value at $\nu=0$, we obtain the imaginary part
of the model susceptibility written as
\begin{equation}
{\rm Im}{\chi}^{* R}_s({\bf Q},\nu) = \frac{c \nu}
{\left[ 1-U {\rm Re}\chi_0^R({\bf Q},0) \right]^2+(c \nu)^2}.
\end{equation}

One can see that
our model susceptibility is formally equivalent to the phenomenological
model susceptibility in the nearly-antiferromagnetic-Fermi-liquid
(NAFL) theory \cite{PINES94,STOJKO98,VILK97b,SPS98}.
This model was introduced by Millis, Monien, and Pines to explain
the NMR experiments in the high-$T_c$ materials \cite{MMP}.
By applying the NAFL theory to the effective models including
the next-nearest-neighbor hopping, anisotropic pseudogap
formation\cite{VILK97b} and band dispersion\cite{SPS98} which
explains the experimental data in the high-$T_c$ materials
have been obtained.

Using our model susceptibility,
the integrand of the imaginary part of the self-energy becomes
\begin{equation}
I^*_{{\bf k}_F}({\bf Q},\omega)
=-\frac{3}{2}U^2
\frac{
  (c/2) \omega \coth \left( \omega/(2T) \right)
}
{
\left[1- U {\rm Re}\chi_0^R({\bf Q},0) \right]^2
+(c \omega)^2
}.
\end{equation}

In Fig.\ \ref{MODELI} (a),
we show the $U$ dependence of ${\rm Im}{\chi}^{* R}_s({\bf Q},\nu)$
for $T/t=0.25$.  The necessary parameters in the model susceptibility
are chosen as $c/t=0.5$ and ${\rm Re}\chi_0({\bf Q},0)=0.4$.
We have roughly evaluated those parameters from the numerical data,
but the accuracy is not important here.
As $U/t$ becomes large, ${\rm Im}{\chi}^{* R}_s({\bf Q},\nu)$
shows a low-energy Stoner enhancement.
In Fig.\ \ref{MODELI} (b) and (c), we show the $U$ evolution of
$I^*_{{\bf k}_F}({\bf Q},\omega)$.
For small $U/t$, we see the $\omega$ dependence similar to
the Fermi-liquid-like $\omega^2$ behavior around $\omega=0$ although
$I^*_{{\bf k}_F}({\bf Q},\omega)$ is the integrand of
the imaginary part of the self-energy.
As $U/t$ increases, this behavior disappears
and finally a very sharp negative peak structure develops
at $\omega=0$.

Thus, even if the density of states does not have a log singularity,
the strong negative peak appears in the integrand
of the imaginary part of the self-energy as the parameter set
approaches the Stoner instability condition.

%%%%%%%%%%%%%%%%%%%%%%%%%%%%%%%%%%%%%%%%%%%%%%%%%%%%%%%%%%%%%%%%%%%%%%%%%%
\subsection{Necessary condition for the pseudogap formation}

As our numerical results have shown, the non-Fermi-liquid behavior
of the self-energy and the pseudogap are related to each other.
In all cases, as we have seen in our results, a pseudogap in
$A({\bf k}_F,\omega)$ accompanies the positive slope of the real
part of the self-energy.
In this subsection, we generally argue the relation between
the self-energy and the spectral function at the Fermi level.
We prove two theorems which hold between the self-energy
and the spectral function.
These theorems determine which conditions in the self-energy
are necessary in order to have a pseudogap in the spectral function.
Through the argument,
we assume that: (i) the imaginary part of the (retarded) self-energy
is always negative and finite, (ii) at $\omega=0$, we have
$(\eps_{{\bf k}_F}-\mu) +{\rm Re}\Sigma^R({\bf k}_F,0) = 0$.
We leave the chemical potential $\mu$ to keep the generality.
Those assumptions are physically reasonable.

Next, we enunciate the two theorems, after that, we give the proofs
of them.

{\it Theorem I} ---
If the imaginary part of the self-energy has a maximum
at the Fermi level,
the spectral function has a maximum at the Fermi level.

One easily sees that the conventional Fermi liquid satisfies
the theorem I.

{\it Theorem II} ---
If the spectral function has a minimum at the Fermi level,
the real part of the self-energy has a positive slope
at the Fermi level.

Note that if the spectral function has a hollow at the Fermi level,
we can say that the spectral function has a two peak structure around
the hollow since the spectral sum over energy is finite.
Hence, to discuss the existence of the pseudogap,
it is sufficient to observe if the spectral function
has a maximum or minimum at the Fermi level.

{\it Proof of theorem I} ---
If $\omega$ is slightly different from $0$, 
from the definition of the spectral function, we have
\begin{equation}
A({\bf k}_F,\omega) = \frac{1}{\pi}
\frac{1}
{
| {\rm Im} \Sigma^R({\bf k}_F,\omega) |
+\left[
   \omega -(\eps_{{\bf k}_F}-\mu) -{\rm Re}\Sigma^R({\bf k}_F,\omega)
\right]^2 / | {\rm Im} \Sigma^R({\bf k}_F,\omega) |
}.
\label{AKW2}
\end{equation}
If $\omega=0$, by the assumption (ii), we obtain
\begin{equation}
A({\bf k}_F,0) = \frac{1}{\pi |{\rm Im} \Sigma^R({\bf k}_F,0)|}.
\end{equation}
From the condition of this theorem,
${\rm Im}\Sigma^R({\bf k}_F,\omega)$ has a maximum at the Fermi level,
and we have
 $| {\rm Im} \Sigma^R({\bf k}_F,\omega) |
> | {\rm Im} \Sigma^R({\bf k}_F,0) |$.
Moreover, the second term of the denominator of Eq.\ (\ref{AKW2}) is
a positive finite, then we find $A({\bf k}_F,0) > A({\bf k}_F,\omega)$.
The theorem I has been proved.

Before going to the proof of theorem II, we give the following lemma
which is used in the proof.

{\it Lemma} ---
If ${\rm Im}\Sigma^R({\bf k}_F,\omega)$ has a maximum (minimum)
at the Fermi level, then ${\rm Re}\Sigma^R({\bf k}_F,\omega)$
has a negative (positive) slope at the Fermi level.
This lemma can be proved using a property of the Kramers-Kronig
relation between ${\rm Im}\Sigma^R({\bf k}_F,\omega)$ and
${\rm Re}\Sigma^R({\bf k}_F,\omega)$.  We give a proof of
this property in the appendix, and the lemma follows immediately
from it.

{\it Proof of theorem II} ---
From Dyson's equation, we obtain that
\begin{equation}
{\rm Re}\Sigma^R({\bf k},\omega) = \omega - (\eps_{\bf k}-\mu)
-\frac{{\rm Re}G^R({\bf k},\omega)}
{ \left[ {\rm Re}G^R({\bf k},\omega) \right]^2
+ \left[ {\rm Im}G^R({\bf k},\omega) \right]^2 }
\label{SG1}
\end{equation}
and
\begin{equation}
{\rm Im}\Sigma^R({\bf k},\omega) =
 \frac{{\rm Im}G^R({\bf k},\omega)}
{ \left[ {\rm Re}G^R({\bf k},\omega) \right]^2
+ \left[ {\rm Im}G^R({\bf k},\omega) \right]^2 }.
\label{SG2}
\end{equation}
At ${\bf k}={\bf k}_F$, the latter equation gives
\begin{equation}
|{\rm Im}\Sigma^R({\bf k}_F,\omega)| =
 \frac{1}
{|{\rm Im}G^R({\bf k}_F,\omega)|
+ \left[ {\rm Re}G^R({\bf k}_F,\omega) \right]^2
/|{\rm Im}G^R({\bf k}_F,\omega)|
}.
\label{IMS}
\end{equation}
At $\omega=0$, from the assumption (ii) and Eq.\ (\ref{SG1}),
we find that ${\rm Re}G^R({\bf k}_F,0)=0$.
Then, from Eq.\ (\ref{IMS}), we have
\begin{equation}
|{\rm Im}\Sigma^R({\bf k}_F,0)| = \frac{1}{|{\rm Im}G^R({\bf k}_F,0)|}.
\end{equation}
Since the spectral function has a minimum at $\omega=0$,
we see $|{\rm Im}G^R({\bf k}_F,0)| < |{\rm Im}G^R({\bf k}_F,\omega)|$
for $\omega$ near zero.  Thus, we have
\begin{equation}
{\rm Im}\Sigma^R({\bf k}_F,0) < {\rm Im}\Sigma^R({\bf k}_F,\omega) < 0,
\end{equation}
i.e., ${\rm Im}\Sigma^R({\bf k}_F,\omega)$ has a minimum at $\omega=0$.
From the lemma, we see that ${\rm Re}\Sigma^R({\bf k}_F,\omega)$ has
a positive slope at $\omega=0$.
The theorem II is finally proved.

We easily see that the reverses of the two theorems do not always hold.
For example, when ${\rm Re}\Sigma^R({\bf k}_F,\omega)$ has a positive
slope, $A({\bf k}_F,\omega)$ may have a single peak at the Fermi level.
However, we can never have a simultaneous maximum
in ${\rm Im}\Sigma^R({\bf k}_F,\omega)$
and a pseudogap in $A({\bf k}_F,\omega)$ at the Fermi level.
The argument we gave here does hold very generally and
it does not depend on the models and the origin of the pseudogap.

The reverse of the lemma is not always true either.
However, if we include an additional condition in
the initial assumption such as
the imaginary part of the self-energy has always
a minimum or a maximum at the Fermi level,
the revere of the lemma is satisfied.
To prove this let's assume this new condition.
From the lemma, ${\rm Re}\Sigma({\bf k}_F,\omega)$
always has a (positive or negative) slope at the Fermi level.
Thus, the maximum (minimum) of ${\rm Im}\Sigma^R({\bf k}_F,\omega)$
is automatically linked to the negative (positive) slope
of ${\rm Re}\Sigma^R({\bf k}_F,\omega)$.
The reverse of the lemma follows from this.

By combining the reverse of the lemma and the theorem I,
we immediately find that
when ${\rm Re}\Sigma^R({\bf k}_F,\omega)$ has a negative slope
at $\omega=0$, the spectral function has a single peak
at the Fermi level.
In other words, as far as the Fermi-liquid-like negative slope
remains in ${\rm Re}\Sigma^R({\bf k}_F,\omega)$ at $\omega=0$,
the pseudogap is never produced in the spectral function
at the Fermi level.

%%%%%%%%%%%%%%%%%%%%%%%%%%%%%%%%%%%%%%%%%%%%%%%%%%%%%%%%%%%%%%%%%%%%%%%%%%
\section{Band dispersion and anisotropic pseudogap}

In Fig.\ \ref{DISP} (a) we show the band dispersion extracted from
the numerical data of the spectral function $A({\bf k}, \omega)$
for several momentum directions.  The dot indicates the peak
of the spectral function.
The peaks roughly follow the original non-interacting
band dispersion except around $(\pi,0)$.
Clear (pseudo)gap structure can be seen around $(\pi,0)$.
In contrast, around $(\pi/2,\pi/2)$, there is no such pseudogap
structure.  This exemplifies the anisotropy of the pseudogap
on the Fermi surface.
We show the structures of the spectral functions
around $(\pi,0)$ in Fig.\ \ref{DISP} (b).
We chose ${\bf k}=(\pi,0)$, $(0.9\pi,0)$, and $(0.85\pi,0)$
along the $k_x$ axis.
By varying ${\bf k}$, the symmetric two peak structure
at ${\bf k}=(\pi,0)$ gradually changes into a single peak
structure with a satellite peak.
The satellite peak corresponds to the shadow band
of the SDW band dispersion.
In our earlier work\cite{TAI97},
another sign of similar shadow band formation around
${\bf k}=(\pi/2,\pi/2)$ and $(\pi,0)$ has been observed
in the dispersion spectrum for a weakly doped Hubbard model.

In Fig.\ \ref{SPCFIG2}
we compare the spectral functions and the self-energies
for ${\bf k}_F=(\pi,0)$ and $(\pi/2,\pi/2)$.
${\rm Re}\Sigma^R({\bf k}_F,\omega)$ at ${\bf k}_F=(\pi,0)$
in Fig.\ \ref{SPCFIG2} (b) has a large positive slope
around $\omega=0$.
${\rm Re}\Sigma^R({\bf k}_F,\omega)$ at ${\bf k}_F=(\pi/2,\pi/2)$
also has a positive slope, but it is still weak.
${\rm Im}\Sigma^R({\bf k}_F,\omega)$ at ${\bf k}_F=(\pi,0)$
in Fig.\ \ref{SPCFIG2} (c) shows a sharp negative peak
at $\omega=0$.  In contrast,
${\rm Im}\Sigma^R({\bf k}_F,\omega)$ at ${\bf k}_F=(\pi/2,\pi/2)$
has a weak minimum at $\omega=0$.
These remarkable momentum dependences of the real and imaginary parts
around $\omega=0$ determines the opening of the pseudogap.

For larger values of $U/t$, we have obtained a pseudogap
in $A({\bf k}, \omega)$ also at ${\bf k}_F=(\pi/2,\pi/2)$.
In Fig. \ \ref{SPC4}, we show the spectral function
for $U/t=3.0$ and at $T/t=0.5$.
We chose the higher temperature to avoid the Stoner instability.
The pseudogap at $(\pi/2,\pi/2)$ is weaker than the one at $(\pi,0)$.
Comparing with the spectral function for $U/t=2.0$ shown
in Fig.\ \ref{SPCFIG2} (a), the intensity of $A({\bf k}_F,\omega)$
in Fig.\ \ref{SPC4} is weak and its width are rather wide.
This indicates band broadening effects due to
the larger value of $U/t$.

To understand the reason for the anisotropy of the pseudogap,
it is sufficient to analyze the imaginary parts of the self-energy
at $\omega=0$ for ${\bf k}_F=(\pi,0)$ and $(\pi/2,\pi/2)$.
There are two factors: the difference of the effective
region of the ${\bf q}$ summation and the different behaviors
of the bosonic excitation energy in the imaginary part
of the self-energy.  The first one is a kind of the selection rule
of the momenta of the fermion-boson excitation and the latter is
the restriction of energy on its excitation.
We explain them as follows.

The distribution function term
$f(\eps_{{\bf k}_F+{\bf q}})+n(\eps_{{\bf k}_F+{\bf q}}-\omega)$
produces an important restriction to the ${\bf q}$ summation
in ${\rm Im}\Sigma({\bf k}_F,0)$.
In Fig.\ \ref{AREA} the approximated summation area
in the $(q_x,q_y)$-plane
is shown by the shaded region in which ${\bf q}$ satisfies the condition
$| f(\eps_{{\bf k}_F+{\bf q}}) + n(\eps_{{\bf k}_F+{\bf q}}) | > 0.5$
for $T/t=0.22$.
We see that for both ${\bf k}_F=(\pi,0)$ and $(\pi/2,\pi/2)$
the region around ${\bf q}={\bf Q}$ contributes
to the ${\bf q}$ summation.
However, the shaded area around ${\bf Q}$ for ${\bf k}_F=(\pi,0)$
is greater than that for $(\pi/2,\pi/2)$.
Thus, ${\rm Im}\Sigma({\bf k}_F,\omega)$ for ${\bf k}_F=(\pi,0)$
has a larger contribution from the spin excitation around
${\bf Q}=(\pi,\pi)$.

The bosonic energy $\eps^b({\bf q}) \equiv
\eps_{{\bf k}_F+{\bf q}}-\omega$ in the expression
of ${\rm Im}\Sigma({\bf k}_F,\omega)$ for ${\bf k}_F=(\pi,0)$
behaves in a totally different way from that for
${\bf k}_F=(\pi/2,\pi/2)$.
To see the behavior for $\omega \sim 0$ and
around ${\bf q} \sim {\bf Q}$, we can write
$\eps^b({\bf Q}+\delta{\bf q})
\sim -\eps_{{\bf k}_F + \delta {\bf q}}$,
where $\delta {\bf q}$ is a small vector to represents
deviation from ${\bf Q}$.
Thus, we see that the behavior of $\eps^b({\bf Q}+\delta{\bf q})$
is determined by the band dispersion $\eps_{\bf k}$ around
${\bf k} = {\bf k}_F$.
Hence, the bosonic energy contribution to ${\rm Im}\Sigma$
for ${\bf k}_F=(\pi,0)$ is quite concentrated within a small region
around $0$ energy
near the saddle point (flat-band) dispersion near $(\pi,0)$.
In contrast, the bosonic energy for ${\bf k}_F=(\pi/2,\pi/2)$
can easily deviate from $0$ as ${\bf q}$ moves
in the summation region
near ${\bf Q}$ since $\eps_{{\bf k}_F}$ has a linear slope
around ${\bf k}_F=(\pi/2,\pi/2)$.
Thus, the low energy enhancement
in the spin fluctuation around ${\bf Q}$ makes
a stronger contribution to ${\rm Im}\Sigma({\bf k}_F,\omega)$
at ${\bf k}_F=(\pi,0)$ than that at ${\bf k}_F=(\pi/2,\pi/2)$.
As a result, the self-energy has a remarkable momentum dependence,
and the anisotropy of the pseudogap is explained.

There are some QMC simulations of the spectral functions
in the weak-coupling regime
at half-filling\cite{CREF95,WHITE92,DUFFY95b}.
With the exception of
the results from Creffield et al. (Ref.\ \onlinecite{CREF95}),
a clear evidence of the anisotropic pseudogap in the spectral function
is lacking in the QMC simulations.
We think that this is also
due to the cluster size effect and the difference in
method in extracting the information of the spectral
function from the Green's function as emphasized in
Ref.\ \onlinecite{CREF95}.
We know of no other example of the anisotropic pseudogap in the
$t$-$U$ Hubbard model.
However, in FLEX calculations, for the doped
Hubbard model\cite{LANGER95,SCHMA96}, a clear momentum dependence
in the imaginary part of the self-energy at $\omega=0$ near
the Fermi momentum was obtained.  They found that 
$|{\rm Im}\Sigma({\bf k}_F,0)|$ has a large value
along the Fermi surface which becomes larger and larger
as ${\bf k}_F$ approaches the $(\pi,0)$ region.
This tendency is consistent with our results at half-filling
shown in Fig.\ \ref{SPCFIG2} (c).

%%%%%%%%%%%%%%%%%%%%%%%%%%%%%%%%%%%%%%%%%%%%%%%%%%%%%%%%%%%%%%%%%%%%%%%%%%
\section{Concluding remarks}
In this work we have studied the pseudogap formation
in the two-dimensional Hubbard model at half-filling.
We obtain a pseudogap formation in the density of states
and in the spectral function.
The pseudogap of the spectral function
is produced when two conditions are satisfied:
(i) a strong anti-ferromagnetic spin fluctuation,
(ii) a nesting condition on the Fermi surface.
The pseudogap on the Fermi energy is highly anisotropic and
its associated symmetry is similar to the $d$-wave symmetry.
The anisotropy is determined by the flatness of the band dispersion.

We emphasize that the pseudogap formation discussed is highly
dependent on the specific character (perfect nesting)
of the free band dispersion or the hopping term in the square lattice.
Thus, the pseudogap is physically
different from a gap formation such as the Mott insulator
transition which takes place in a strong coupling regime.
Our pseudogap can take place
even in the weak coupling regime.

We have focused on the undoped single-band Hubbard model with
nearest neighbor hopping.
Unfortunately, the model we have used may be quite simple
to make a quantitative comparison with the experimental data.
Nevertheless, it is possible to apply our argument in the pseudogap
originated, for example, in the Hubbard model with
the next nearest neighbor hopping ($t$-$t'$-$U$ model).  In this model,
the nesting condition
associated with the momentum ${\bf Q}$ holds only around the $(\pi,0)$
region (and, of course, also at the other symmetric three parts in
the Brillouin zone).  Hence, the pseudogap will open in this region.
However, the region in the $k_x= \pm k_y$ direction on the Fermi surface
does not satisfy the nesting condition.  Thus, we can predict
that the electronic states around this diagonal region will remain
in the Fermi liquid regime.
Our argument is consistent with the physical picture of
the so-called {\it hot} and {\it cold} quasiparticles
by B. Stojkovi\'c and Pines\cite{STOJKO98}.
Investigation with more realistic models will be done
in future works.

Very recently, the information of the self-energy has been extracted
from the angle-resolved photoemission data\cite{NORMAN98B}.
In the work, it has been obtained that the observed normal state
pseudogap accompanies a sharp negative peak in ${\rm Im}\Sigma^R$
and a positive slope in ${\rm Re}\Sigma^R$.
The tendency of their data is qualitatively in good agreement
with our results.

In the present work we have used a paramagnon-theory self-energy
to calculate the electronic states.  This treatment has
a restriction due to the Stoner criterion
in the RPA susceptibility.
However, we believe that
already in the level of this scheme, it contains
important ingredients for the pseudogap formation and
the partial destruction of the Fermi-liquid quasiparticles
at the Fermi level.
In particular, the anisotropic pseudogap formation in the spectral
function in the calculated band dispersion indicates
the coexistence of the Fermi-liquid-like quasiparticles
and the SDW-like quasiparticles at the Fermi level.

%%%%%%%%%%%%%%%%%%%%%%%%%%%%%%%%%%%%%%%%%%%%%%%%%%%%%%%%%%%%%%%%%%%%%%%
\acknowledgements
We would like to acknowledge useful discussions with
S. L. Garavelli and P. E. de Brito.
This work was supported by the Conselho Nacional de
Desenvolvimento Cient\'{\i}fico e Tecnol\'ogico - CNPq and
by the Financiadora de Estudos e Projetos - FINEP.
Most part of the numerical calculations of this work
have been done with the supercomputing system
at the Institute for Materials Research, Tohoku University, Japan.
T.S. thanks the Material Science Group at IMR Tohoku University
for the use of the supercomputing facilities.

%%%%%%%%%%%%%%%%%%%%%%%%%%%%%%%%%%%%%%%%%%%%%%%%%%%%%%%%%%%%%%%%%%%%%%%%%%%
\appendix

%%%%%%%%%%%%%%%%%%%%%%%%%%%%%%%%%%%%%%%%%%%%%%%%%%%%%%%%%%%%%%
\section*{A relation in the Kramers-Kronig transformation}

In this appendix we give a simple proof of a relation
obtained from the Kramers-Kronig transformation.
We apply this relation to the real and imaginary parts
of the self-energy in the main text of the present paper.
The relation holds for any smooth functions.
Suppose the following Kramers-Kronig relation
for such two functions ${\rm g}(x)$ and ${\rm h}(x)$,
\begin{equation}
{\rm g}(x) = \frac{P}{\pi} \int_{-\infty}^{\infty} dx'
\frac{{\rm h}(x')}{x'-x}.
\end{equation}
Then, the relation can be mentioned as follows.
If ${\rm h}(x)$ has
a maximum (minimum) at a certain point $x=x_0$, ${\rm g}(x)$ has
a positive (negative) slope at $x_0$.
We assume that ${\rm h}(x)$ has a peak at $x=x_0$
and it can be expanded around $x_0$ as ${\rm h}(x) \sim
{\rm h}(x_0)+(\gamma_{\rm h}/2)(x-x_0)^2$.
This is reasonable as far as ${\rm h}(x)$ is a smooth function.
Here, $\gamma_{\rm h}$ is the second derivative of ${\rm h}(x)$
at $x=x_0$.
If $\gamma_{\rm h}>0$ $(<0)$, then the peak is a minimum (maximum).

Let's observe the difference between ${\rm g}(x_0+\eps)$ and
${\rm g}(x_0)$ with $\eps$ being a small shift from $x_0$.
We obtain that
\begin{equation}
{\rm g}(x_0+\eps) - {\rm g}(x_0)
= \frac{P}{\pi} \int_{-\infty}^{\infty} dx'
\frac{\eps {\rm h}(x')}
{(x'-x_0)^2-\eps (x'-x_0)}.
\end{equation}
We easily see that the function $\eps/[(x'-x_0)^2-\eps (x'-x_0)]$
decreases rapidly as $x'$ deviates from $x_0$ and $x_0+\eps$.
We can approximate the integration using the expansion
of ${\rm h}(x)$ around $x_0$ as
\begin{equation}
{\rm g}(x_0+\eps) - {\rm g}(x_0)
\sim \frac{P}{\pi} \int_{x_1}^{x_2} dx'
\frac{\eps \left[{\rm h}(x_0)+(\gamma_{\rm h}/2)(x'-x_0)^2\right]}
{(x'-x_0)^2-\eps (x'-x_0)}.
\end{equation}
where $x_1 < x_0 < x_2$, and besides,
$|x_2-x_0|$, $|x_1-x_0| \gg |\eps|$.
We have chosen $x_1$ and $x_2$ in a way which
the expansion of ${\rm h}(x)$ is valid.
By evaluating the integral up to the leading order of $\eps$,
we obtain
\begin{equation}
\frac{ {\rm g}(x_0+\eps) - {\rm g}(x_0) }{\eps} \sim
\frac{(x_2-x_1) \gamma_{\rm h}}{2 \pi}.
\end{equation}
Thus, for $\gamma_{\rm h}>0$ ($\gamma_{\rm h}<0$) at $x=x_0$,
the slope of ${\rm g}(x_0)$ is positive (negative).

By applying the same argument to the opposite transformation
defined as
\begin{equation}
{\rm h}(x) = -\frac{P}{\pi} \int_{-\infty}^{\infty} dx'
\frac{{\rm g}(x')}{x'-x},
\end{equation}
we find that if ${\rm g}(x)$ has a maximum (minimum) at $x=x_0$,
the slope of ${\rm h}(x)$ at $x=x_0$ is negative (positive).

In the above derivation, we assumed that the functions of our
interests can be
approximated with a quadratic form around a peak point we
have interests.
One can show the same relation in the case that
the function behaves as ${\rm h}(x) \propto |x-x_0|$ around $x=x_0$.
This is what happens
if ${\rm h}(x)$ is the imaginary part of the self-energy
of a marginal Fermi liquid\cite{MFL}.
Even if the peak is a $\delta$-function, 
the relation we argued is still applicable
although the slope becomes infinite.  One example of this can be seen
in the mean-field Green's function for the spin-density-wave states.

%%%%%%%%%%%%%%%%%%%%%%%%%%%%%%%%%%%%%%%%%%%

%%%%%

\begin{figure}[h]
\caption{
$U$ dependence of the density of states $N(\omega)$
at $T/t=0.5$.
} % end of \caption
\label{DOS1}
\end{figure}

\begin{figure}[h]
\caption{
Temperature dependence of the density of states $N(\omega)$
for $U/t=2.0$.  Inset shows the overall structures of $N(\omega)$.
} % end of \caption
\label{DOS2}
\end{figure}

\begin{figure}[h]
\caption{
Temperature evolution of
(a) the spectral function and (b) the real and (c) imaginary
parts of the self-energy for $U/t=0.2$ and at ${\bf k}_F=(\pi,0)$
} % end of \caption
\label{SPCFIG}
\end{figure}

\begin{figure}[h]
\caption{
Temperature dependence of (a) ${\rm Im}\chi_s({\bf Q},\nu)$
and (b) the integrand $I_{{\bf k}_F}({\bf Q},\omega)$ of the imaginary
part of the self-energy for $U/t=2.0$.
} % end of \caption
\label{INTGRD}
\end{figure}

\begin{figure}[h]
\caption{
$U$ dependence of the excitation spectra
of (a) the model susceptibility (see the text)
${\rm Im}\chi_s^*({\bf Q},\nu)$ and
(b), (c) the integrand $I^*_{{\bf k}_F}({\bf Q},\omega)$
of the imaginary part of the self-energy for the model susceptibility.
} % end of \caption
\label{MODELI}
\end{figure}

\begin{figure}[h]
\caption{
(a) Band dispersion for several momentum directions
and (b) $A({\bf k},\omega)$ around $(\pi,0)$ region
calculated for $U/t=2.0$ and at $T/t=0.22$.
} % end of \caption
\label{DISP}
\end{figure}

\begin{figure}[h]
\caption{
(a) Spectral function at ${\bf k}_F=(\pi,0)$ (solid line)
and $(\pi/2,\pi/2)$ (dotted line),
and the corresponding (b) real and (c) imaginary parts
of the self-energy.
The parameters are $U/t=2.0$ and $T/t=0.22$.
} % end of \caption
\label{SPCFIG2}
\end{figure}

\begin{figure}[h]
\caption{
Spectral function at ${\bf k}_F=(\pi,0)$ (solid line)
and $(\pi/2,\pi/2)$ (dotted line) for $U/t=3.0$ and at $T/t=0.5$.
Note the differences in scale compared with Fig.\ 7 (a).
} % end of \caption
\label{SPC4}
\end{figure}

\begin{figure}[h]
\caption{
Main contribution area (shown with shade)
to the ${\bf q}$ summation in ${\rm Im}\Sigma({\bf k}_F,0)$
for $T/t=0.22$ and at (a) ${\bf k}_F = (\pi/2,\pi/2)$ and
(b) ${\bf k}_F = (\pi,0)$ .
Momenta ${\bf q}$'s in the region satisfy
$| f(\eps_{{\bf k}_F+{\bf q}}) + n(\eps_{{\bf k}_F+{\bf q}}) |
> 0.5$.
} % end of \caption
\label{AREA}
\end{figure}


\begin{thebibliography}{99}
\bibitem{PI0}
Hereafter, "$(\pi,0)$" represents all symmetric momentum points
on the non-interacting Fermi surface,
i.e., $(\pm \pi,0)$, $(0, \pm \pi)$.
We also use "$(\pi/2,\pi/2)$" in the same manner.
\end{thebibliography}
\end{document}